\documentclass{acm_proc_article-sp}
\pdfoutput=1

\usepackage{color}
\usepackage{booktabs} % This one is for nice tables
\usepackage{multirow} % This is for columns that span multiple rows in tables
\usepackage{algorithmic} % These two to render algorithms nicely
\usepackage{algorithm}
\usepackage{verbatim} % For multi-line comments
\usepackage[normalem]{ulem} % Used here mainly for struck-out text
\usepackage{url}  % This allows to use \url{...} construct
\usepackage[pdfdisplaydoctitle=true,
colorlinks=true,
citecolor=blue,
linkcolor=black,
urlcolor=black]{hyperref}
\graphicspath{{images/}}

\makeatletter
\let\@copyrightspace\relax
\makeatother

\begin{document}

% BLUE COLOUR
%{\color{blue}
%} % END OF BLUE COLOUR

% That's how you display math formulae on their own lines:
% \begin{displaymath}E=\lceil N \rceil \end{displaymath}

% Set author information, subject and keywords
\hypersetup{pdftitle=Automated Design of Two-Layer Fat-Tree Networks}
\hypersetup{pdfauthor=Konstantin S. Solnushkin}
\hypersetup{pdfsubject=Network Design}
\hypersetup{pdfkeywords=Fat-tree network; Data Centre; Supercomputer; Cluster; Automated design}
% Red links, pointing to different objects inside the document, are a bit loud-coloured. We make them white and invisible:
% We used to have borders around links to bibliography:
%\hypersetup{colorlinks,linkbordercolor=1 1 1}

% Introduce a new environment, "Algorithm":
% XXX: Superceded by package "algorithm":
%\newdef{algorithm}{Algorithm}

\title{Automated Design of Two-Layer Fat-Tree Networks}

\numberofauthors{1}

\author{
\alignauthor
Konstantin S. Solnushkin\\
       \email{konstantin@solnushkin.org}
}

\maketitle

\begin{abstract}
This paper presents an algorithm to automatically design two-level fat-tree networks, such as ones widely used in large-scale data centres and cluster supercomputers. The two levels may each use a different type of switches from design database to achieve an optimal network structure. Links between layers can run in bundles to simplify cabling. Several sample network designs are examined and their technical and economic characteristics are discussed.

The characteristic feature of our approach is that real life equipment prices and values of technical characteristics are used. This allows to select an optimal combination of hardware to build the network (including semi-populated configurations of modular switches) and accurately estimate the cost of this network. We also show how technical characteristics of the network can be derived from its per-port metrics and suggest heuristics for equipment placement.

The algorithm is useful as a part of a bigger design procedure that selects optimal hardware of cluster supercomputer as a whole. Therefore the article is focused on the use of fat-trees for high-performance computing, although the results are valid for any type of data centres.
\end{abstract}

% Categories, as per: http://www.acm.org/about/class/ccs98-html and http://www.acm.org/about/class/how-to-use

% General terms:

% Keywords:

% Introduction
\section{Introduction}
Parallel computers use many types of networks to interconnect its computing elements. Frequently used topologies include stars, meshes, tori and trees.

``Beowulf''-style cluster supercomputers often  employ fat-tree topologies built using readily available off-the-shelf InfiniBand hardware. We describe an algorithm that allows to automatically design fat-tree networks with a variety of objective functions, with the most obvious example being the total cost of network. The algorithm is implemented in a software tool~\cite{solnushkin-network-design-tool}.

This algorithm is intended to be used as a part of a CAD system for cluster supercomputers~\cite{solnushkin2012computer}. Such a system would iterate through different combinations of hardware, varying the number of compute nodes and other parameters. Thus, designing an interconnection network for every hardware combination under review is a self-contained and highly repetitive operation that must be performed efficiently.

Many researchers of fat-tree networks concentrate on general properties of such networks and big fabrics that could be built using them. We focus on real-life scenarios, tailoring network designs to the number of network endpoints and available switches. For example, our approach allows to select optimal configurations of modular switches, with just the right number of leaf modules installed.

Current ASIC technology enabled the appearance of readily available, off-the-shelf InfiniBand switches with $P=36$ ports. This allows to build two-level fat-trees with as much as $P^2/2=648$ cluster nodes. For many typical installations this is enough.

However, vendors also provide high-radix modular switches, which internally implement a two-level fat-tree. Switches with up to $P=648$ ports (in non-blocking configurations) are available, hence networks with more than 200K nodes can be built with the proposed algorithm -- this far exceeds the demands of even the most powerful today's supercomputers.

On the other hand, intermediate-sized designs that do not use the full capacity provided by switches tend to have unused ports unless designed carefully. If no network expansion is anticipated, unused ports represent a waste of hardware resources. Therefore our algorithm tries to minimize the number of unused ports. Additionally, the algorithm reports if links between switches can run in bundles. Such bundles can be implemented with cables that aggregate multiple links, e.g., 12x instead of 4x InfiniBand cables. This results in a lower number of cables and reduced cable bulk.

To obtain cable costs, equipment placement and routing can be performed (not discussed in this article), or alternatively an ``average'' cable price can be used, see section \ref{fat-tree:calculating-cable-costs} for details.

During the design process, other characteristics of interconnection networks, such as reliability, can be estimated and used as design constraints or as a part of a complex objective function.

The rest of the article is organized as follows. Section \ref{fat-tree:related-work} describes existing work in the field of fat-tree networks and their economic issues. Section \ref{fat-tree:algorithm} introduces the main algorithm, and section \ref{fat-tree:discussion} discusses it. In section \ref{fat-tree:experimental-results} we conduct a sample run of the algorithm and present the results. Section \ref{fat-tree:other-design-considerations} explains how to obtain technical and economic characteristics of fat-trees using per-port metrics, discusses design for future expansion and proposes heuristics for equipment placement. Finally, section \ref{fat-tree:conclusions} concludes the article.

% Related work
\section{Related work}
\label{fat-tree:related-work}

Fat-trees were initially introduced by C.~Leiserson \cite{leiserson1985fattrees}. The mathematical formalism to describe their structure, ``k-ary n-trees'', was proposed by Petrini and Vanneschi \cite{petrini1997kary}. Zahavi \cite{zahavi2012fattree} further introduces two other formalisms for describing fat-trees, \textit{Parallel Ports Generalized Fat-Trees}, where links between switches can run in parallel, and \textit{Real Life Fat-Trees} where bandwidth between layers stays constant to guarantee content-free operation.

A tool called \textit{NetWires} \cite{aggregate.org:netwires} was created by H.G.Dietz as part of the bigger project \textit{Cluster Design Rules} \cite{dieter2005automatic}. \textit{Netwires} is able to design different types of interconnection networks, including trees, tori and a specific \textit{Flat Neighbourhood Network}, using user-supplied parameters, and outputs a wiring diagram. Aside from the number of required switches, no other technical or economic characteristics are assessed. Our approach is different in that we only require a few input parameters from the user, and iterate through other parameters automatically, trying to find a combination that yields an optimal value to a certain objective function subject to constraints.

Gupta and Dally \cite{gupta2006topology} suggested a tool to optimize network topology, in the broad class of hybrid Clos-torus networks. Cost, packaging and performance constraints can be specified. This tool is most valuable for building custom interconnection solutions when arbitrary topologies are feasible, contrary to the case of using commodity switches where most parameters are fixed but optimization can take actual prices into account.

Al-Fares \textit{et al.} \cite{al2008scalable} proposed to use fat-trees for generic data centre networks, using commodity hardware. Farrington \textit{et al.} \cite{farrington2009data} followed up, suggesting to build a 3,456-port data centre switch with commodity chips (``merchant silicon'') internally connected in a fat-tree topology. They also advice to use optical fibre cables with as much as 72 or even 120 separate fibres (strands) to minimize the volume and weight of cable bundles for inter-switch links.

Mudigonda \textit{et al.} \cite{mudigonda2011taming} introduced \textit{Perseus}, a framework to design fat-tree and HyperX topologies for data centres, and elaborated on cable tracing issues. However, fat-tree topologies built by \textit{Perseus} use identical switches on all levels.

Parallel applications typically exhibit locality of communications. Therefore in multi-level non-blocking fat-tree networks the bandwidth offered by upper levels may remain underutilized. On intermediate levels, switch ports can be redistributed so that the number of links to the lower level is bigger than to the upper level. This reduces ``fatness'' of a tree, providing substantial hardware savings in terms of switches and links.

Navaridas \textit{et al.} \cite{navaridas2010reducing} introduced such a reduced topology, \textit{thin-tree}, and analysed its behaviour using simulation and several synthetic workloads. Overall, for the mix of workloads, different configurations of the reduced topology were found beneficial in terms of ``performance/cost'' ratio compared to traditional fat-trees, especially when collective operations were only lightly used. They add, however, that in the absence of a topology-aware scheduler, neighbouring processes may be assigned to physically distant processing nodes, requiring full bandwidth at upper levels and thus rendering reduced topologies useless.

Kamil \textit{et al.} \cite{kamil2010communication} similarly proposed a reduced topology, but used communication patterns of actual parallel applications for analysis.

Kim \textit{et al.} \cite{kim2007flattened} introduced a flattened butterfly topology, providing detailed analysis of cost breakdown for electrical cables. 12x InfiniBand cables, aggregating three 4x links, were shown to be more economical than separate 4x cables and to additionally reduce cable bulk. Their subsequent work \cite{kim2008technology} compared cost models for electrical and active optical cables, showing that in 2008 prices, optical cables are less expensive starting from 10m. Parker and Scott \cite{parker2009impact} further advocate for the adoption of optical interconnects.

Singla \textit{et al.} \cite{singla2012jellyfish} proposed to abstain from rigid network structures such as fat-trees, and connect switches in a random order, in a topology called \textit{Jellyfish}. They found that with the same performance figures and the same network equipment as the fat-tree, their topology supports more servers (performance results were obtained via simulation with random permutation traffic). Another benefit of \textit{Jellyfish} is the ability of incremental expansion.

% Algorithm
\section{Algorithm}
\label{fat-tree:algorithm}

Let us consider the algorithm to design fat-tree networks with two levels of switches (namely, \textit{edge} and \textit{core} layers). Suppose we have two databases, for edge and core switches, respectively, with each switch characterized primarily by the number of its ports. For the number of ports of a specific edge switch we will use the designation $P_E$, and for a core switch we will use $P_C$.

Some switch models can be used for building both core and edge levels, and can be present in both databases. The two layers of network can employ different types of switches, but switches within the same layer are identical.

Let $\mathbb{E}$ and $\mathbb{C}$ be the sets of edge and core switches. Each $i$-th switch is characterized by its model and the number of ports, e.g.: $\mathbb{C}=\{\langle M_{C_i}, P_{C_i} \rangle\}$. These sets are the algorithm's input. Their structure allows them to contain several models of switches with the same number of ports but with differing characteristics, such as cost, reliability, energy consumption, etc.

For blade servers, which are installed into enclosures, edge-level switches are also installed in the same enclosures, and thus $\mathbb{E}$ usually contains only one element -- a single switch, compatible with the enclosure. Ordinary rack-mounted servers can, on the contrary, use a variety of edge-level switches.

Apart from $\mathbb{E}$ and $\mathbb{C}$, other inputs for the algorithm are $N$, the number of compute nodes that need to be interconnected, and $Bl$, the blocking (oversubscription) factor, which denotes the decrease in bandwidth available to compute nodes compared to a full, non-blocking fat-tree.

The outputs of the algorithm are models of edge and core switches used to obtain the optimal design, as well as $E$ and $C$, the number of edge and core switches, respectively, and $f$, the value of the objective function.

% START of algorithm
\renewcommand{\algorithmicrequire}{\textbf{Input:}}
\renewcommand{\algorithmicensure}{\textbf{Goal:}}

\begin{algorithm}
\caption{Design a two-level fat-tree network}
\label{fat-tree:alg-fat-tree}
\begin{algorithmic}[1]
\REQUIRE ~ \\
$N$: Number of nodes to interconnect \\
$Bl$: Blocking factor \\
$\mathbb{E}, \mathbb{C}$: Sets of edge and core switches
\ENSURE Optimal network structure: \\
$E, C$: Number of edge and core switches \\
$Bl_r$: Resulting blocking factor \\
$B$: Number of links in a bundle \\
$L$: Number of cables \\
$f$: Objective function for the optimal network structure

% Trivial case of two enclosures
\STATE \COMMENT{ First trivial case: }
\IF{(Using blade servers) \textbf{and} (Only two enclosures)}
\label{fat-tree:alg-fat-tree:trivial-1}
  \STATE Trivial case 1: connect enclosures with cables
  \STATE Compute $f_1$
\ENDIF

% Trivial case of star network
\STATE \COMMENT{ Second trivial case: }
\IF{$\exists \langle M, P \rangle \in \mathbb{E \cup C}: P \ge N$}
\label{fat-tree:alg-fat-tree:trivial-2}
  \STATE \COMMENT{ If there exists a switch with $N$ or more ports }
  \STATE Trivial case 2: use star network
  \STATE Compute $f_2$
\ENDIF

% Start of outer loop
\STATE \COMMENT{ Main loop: iterate through edge switches }
\FORALL{edge switches $\langle M_{E_i}, P_{E_i} \rangle \in \mathbb{E}$}

	% Compute number of edge switches
	\STATE $P_{En_i} \leftarrow \lfloor P_{E_i} \cdot (Bl / (1 + Bl))\rfloor$ \COMMENT{ Ports to nodes }
	\label{fat-tree:alg-fat-tree:ports-to-nodes}
	\STATE $P_{Ec_i} \leftarrow P_{E_i} - P_{En_i}$ \COMMENT{ Ports to core level }
	\label{fat-tree:alg-fat-tree:ports-to-core}
	\STATE $Bl_r \leftarrow P_{En_i} / P_{Ec_i}$ \COMMENT{ Resulting blocking }
	\label{fat-tree:alg-fat-tree:resulting-blocking}
	\STATE $E_i \leftarrow \lceil N / P_{En_i} \rceil$ \COMMENT{ Number of edge switches }
\label{fat-tree:alg-fat-tree:calculate-edge}
	
	\FORALL{core switches $\langle M_{C_j}, P_{C_j} \rangle \in \mathbb{C}$}
	\label{fat-tree:alg-fat-tree:iterate-core}

		\IF[ Core switch suitable ]{$P_{C_j} \ge E_i$}
		\label{fat-tree:alg-fat-tree:check-port-count}
			\STATE \COMMENT{ Try core switch $M_{C_j}$ }
			\STATE $B \leftarrow$ min~($P_{C_j} \div E, P_{Ec_i}$) \COMMENT{ Links in a bundle }
	\label{fat-tree:alg-fat-tree:links-in-bundle}
			\STATE $C \leftarrow \lceil P_{Ec_i} / B \rceil$ \COMMENT{ Number of core switches }
			\STATE $L \leftarrow N + E_i \cdot P_{Ec_i}$ \COMMENT{ Number of cables }
			\STATE Compute $f_{i,j}=f(E_i, C_j)$
		\ENDIF
	\ENDFOR

\ENDFOR % End of outer loop

\STATE Choose optimal combination of $M_C$ and $M_E$: $f_3=min~f_{i,j}$

\STATE Output optimal network structure: $f=min(f_1,f_2,f_3)$
\label{fat-tree:alg-fat-tree:globally-optimal}

\end{algorithmic}
\end{algorithm}
% END of algorithm

In a fat-tree network with a blocking factor $Bl$ and edge switches with $P_E$ ports, $P_{En} = \lfloor P_E \cdot (Bl / (1 + Bl))\rfloor$ of those ports are used to connect compute nodes, and the rest are used to connect the edge switch to the core layer. Under these conditions, in order to connect all $N$ nodes, $E = \lceil N / P_{En} \rceil$ edge switches are required.

The remaining ports on edge switches are connected to core layer switches. When building the core layer, each port on an edge switch is connected to a different core switch. It means that a core switch must have at least as many ports as there are edge switches. For example, first ports of all edge switches will connect to the same core switch. As a result, core switches must have at least $P_{Cmin} \ge E$ ports.

Similarly, if a core switch has $P_C \ge P_{Cmin}$ ports, it can be connected to a maximum of $P_C$ edge switches, each of those having $P_{En}$ compute nodes connected to it. Therefore, the maximum number of nodes that could be connected is $N_{max}=P_C \cdot P_{En}$.

The algorithm structure is as follows. First we check for two trivial cases where a full two-level fat-tree network is not required. Then we iterate through a set of edge switches. For every edge switch in a set, we evaluate multiple possible network designs, trying suitable core switches, and choose one of them. Finally, the best design over all iterations is selected.

% START of algorithm description
Let us describe the algorithm by stages.
\begin{enumerate}
\label{fat-tree:alg-fat-tree-description}

\item \label{fat-tree:alg-fat-tree-description-trivial1} The first stage is to check for the trivial case of two blade enclosures (line~\ref{fat-tree:alg-fat-tree:trivial-1}). In this set-up, there are two enclosures of $P_E/2$ servers, each fitted with built-in edge switches with $P_E$ ports. Half of the ports of each switch are connected to servers. The two switches can be directly connected together with $P_E/2$ cables, and a core level is not necessary.

A cheaper alternative is to replace one of the switches with a ``pass through'' panel, also allowing to directly connect this enclosure's servers to the remaining second switch.

(In case of rack-mounted servers, these complications are irrelevant, because two blocks of $P_E/2$ servers can always be connected with a single edge switch with $P_E$ ports).

We check if this configuration satisfies design constraints, and if yes, compute the value $f_1$ of the objective function (in particular, expandability constraints could be violated).

\item \label{fat-tree:alg-fat-tree-description-trivial2}  The second stage concerns the trivial case of a star network (line~\ref{fat-tree:alg-fat-tree:trivial-2}). If there exists a switch, in either $\mathbb{E}$ or $\mathbb{C}$, with enough ports to accommodate all $N$ nodes, it can be used to build a star network. If several such switches exist, we choose one. Similar to the above case, the value of the objective function, $f_2$, is then computed.

\item \label{fat-tree:alg-fat-tree-description-main-loop} The main loop iterates over available edge switches using index $i$.

\begin{enumerate} % Loop for edge switches
\item For every switch model, we calculate: $P_{En_i}$, the number of ports that are connected to compute nodes (line \ref{fat-tree:alg-fat-tree:ports-to-nodes}), $P_{Ec_i}$, the number of ports connected to the core level (line \ref{fat-tree:alg-fat-tree:ports-to-core}), $Bl_r$, resulting blocking factor (line \ref{fat-tree:alg-fat-tree:resulting-blocking}), and finally $E_i$, the number of required edge switches (line~\ref{fat-tree:alg-fat-tree:calculate-edge}).

\item We then iterate through all core switches using index $j$ (line \ref{fat-tree:alg-fat-tree:iterate-core}). If the number of ports on the core switch makes it suitable, we perform the following actions.
% The innermost loop
\begin{enumerate}
\item Calculate $B$, the number of links that run in parallel between edge and core switches (line \ref{fat-tree:alg-fat-tree:links-in-bundle}).

The core level is built in the following way. We take one core switch. For every edge switch, we connect its first port to the core switch. As we have $E$ edge switches, this operation will occupy $E$ ports on the core switch.

Now, we repeat this step several times until we run out of ports on the core switch. If this step is performed for a total of $B$ times, then each of the edge switches becomes connected to the core switch with a bundle of $B$ links. $B$ can be obtained with a simple equation: $B=P_{C_j} \div E$.

In certain rare cases with high blocking factors (see Example 3 below), only one core switch is necessary to connect together all edge switches, and then links from all ports on the edge switch directed towards the core level form a single bundle $B=P_{Ec_j}$. Line \ref{fat-tree:alg-fat-tree:links-in-bundle} handles this scenario using the $min$ function.

\item After determining $B$, we calculate the number of core switches $C$.

\item At this point, the number of edge and core switches becomes known, hence we can calculate the value of the objective function $f_{i,j}$ for this particular fat-tree configuration.

\end{enumerate}
% END of the innermost loop

\item We choose the optimal fat-tree configuration: $f_3=min~f_{i,j}$ (or, alternatively, present a human designer with several choices)

\end{enumerate}

\item From all combinations obtained with the previous steps (trivial cases \ref{fat-tree:alg-fat-tree-description-trivial1}, \ref{fat-tree:alg-fat-tree-description-trivial2}) and main loop (\ref{fat-tree:alg-fat-tree-description-main-loop}), we choose the one with the optimal value of the objective function (line~\ref{fat-tree:alg-fat-tree:globally-optimal}).

\end{enumerate}

% END of algorithm description

\begin{figure*}
\centering
\includegraphics[width=120mm,height=40mm]{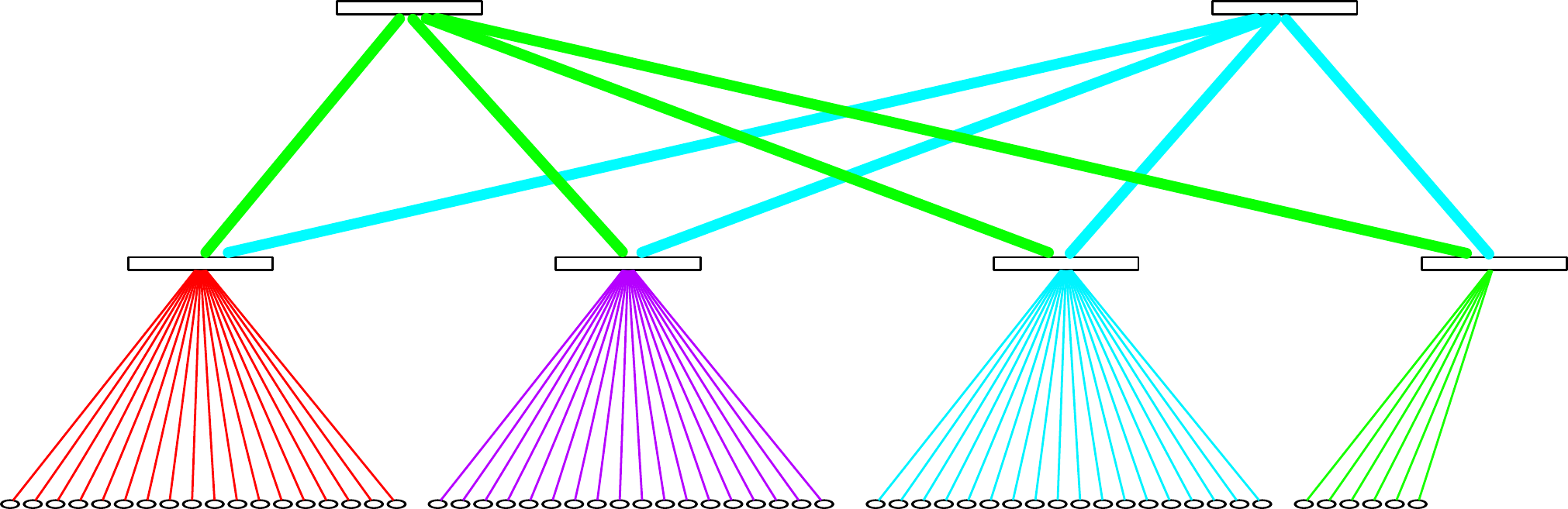}
\caption{Network of $N=60$ nodes, created by the proposed algorithm. Thick lines between edge and core levels represent bundles of nine network links.}
\label{fat-tree:fig:60_nodes_01}
\end{figure*}

Example 1. Suppose we need to interconnect $N=60$ nodes using 36-port switches ($P_E=P_C=36$) with a non-blocking network ($Bl=1$). The algorithm would return $E=4$ edge switches, $C=2$ core switches, and $B=9$ links in a bundle.

The wiring diagram for the resultant network is shown on Figure~\ref{fat-tree:fig:60_nodes_01}. Note that on the rightmost edge switch only 6 ports are utilized, and 12 ports are left unused. Lines between edge and core switches are thicker to represent multiple links connecting switches in edge and core layers. In this case, bundles of $B=9$ links are used. Running links in bundles allows for greater maintainability. Additionally, using 12x InfiniBand cables that aggregate three 4x links allows to decrease the number of physical cables in a bundle to only three.

Example 2. Let us design a network for $N=1200$ nodes, using a blocking factor of $Bl=2$. We will use edge switches with $P_E=36$ ports and core switches with $P_C=108$ ports. Out of 36 ports on the edge switch, $P_{En}=24$ will be connected to compute nodes, and the remaining $P_{Ec}=12$ ports will be connected to the core layer. This provides the blocking factor of $Bl=24/12=2$. The algorithm would return $E=50$ edge switches, $C=6$ core switches, and $B=2$ links in a bundle.

Example 3. Let us now design a network for $N=280$ nodes with an artificially high blocking factor $Bl=11$, using 36-port switches (such a configuration is unsuitable for HPC workloads, and is more relevant for generic data centre environments). The algorithm would distribute ports on edge switches in the following way: $P_{En}=33$ ports will be connected to compute nodes, and $P_{Ec}=36-33=3$ ports will be connected to the core level. The resulting blocking factor is $Bl_r=33/3=11$. The number of edge switches is $E=9$. Only three ports on edge switches are available for connecting to the core level, therefore they will form a single bundle: $B=3$. The number of core switches is then $C=1$.

Note that, when the number of edge switches $E$ is determined, there are two possible scenarios of connecting compute nodes to edge switches: (1) connect as many nodes to each switch as possible, and leave the last switch underutilized (see Fig.~\ref{fat-tree:fig:60_nodes_01}), or (2) distribute compute nodes uniformly between all switches. The latter scenario can, in very rare cases, lead to a lower number of required core switches. However, it also leads to difficulties when expanding the network, because there may not be enough free space in racks. Our software tool \cite{solnushkin-network-design-tool} that implements the algorithm actually checks for that condition, and uses this scenario only if it provides hardware savings \textit{and} if the user didn't specify preference for expandability.

% Discussion
\section{Discussion}
\label{fat-tree:discussion}

\subsection{Switch configurations}
\label{fat-tree:switch-configurations}
We consider two types of switches. First is an ordinary commodity non-modular switch with some redundant components. An example would be a typical off-the-shelf 36-port InfiniBand FDR switch, with redundant fans and an optionally redundant power supply, but with a non-redundant management board.

The second type is the modular switch. Consider, for example, a 144-port InfiniBand switch, equipped with 9 line cards, each allowing to connect 16 nodes in a non-blocking configuration. Fabric boards are used to provide internal fabric of the switch, and all four fabric boards must be installed to make the switch non-blocking. Line cards and fabric boards are installed into the chassis, which also contains redundant power supplies, redundant fans, and redundant management boards.

(Such a switch itself contains a two-level fat-tree, with links between core and edge levels running in bundles of $B=4$ and implemented as traces on its backplane printed circuit board. Parameters of this fat-tree network are as follows: $P_E=32$, $P_C=36$, $E=9$, $C=4$).

Modular switches have large port counts, this reduces the overall number of switches in the network, improving manageability and simplifying cable routing. Additionally, they allow for future expandability by adding more line cards when required. These benefits often outweigh their higher prices per port.

When the number of nodes to be interconnected is lower than the number of ports provided by the fully configured modular switch, a reduced configuration can be used, with fewer line cards installed. This allows to significantly decrease the cost of the switch compared to the full configuration.

Different reduced configurations are treated as different models of switches in the databases $\mathbb{E}$ and $\mathbb{C}$, because they have differing technical and economic characteristics.

\subsection{Design constraints and objective functions}
Objective functions can be diverse, and various design constraints can be specified. Let us confine ourselves to a single example. Suppose we need to choose one of the following switches for the core level: (a) 144-port fully configured modular switch, or (b) partially configured 324-port modular switch, with 144 configured ports. The former takes up less space in a rack, while the latter provides for future expansion.

If constraints on equipment size are imposed, the 144-port switch will be used, and the 324-port switch might not be even tried if it violates constraints. Conversely, if constraints on future expandability are imposed, the 144-port switch might get discarded. If no constraints are imposed, exhaustive search will be performed: the value of the objective function (e.g., the total cost of ownership of the network) will be calculated for both variants to make the decision.

\subsection{Cable count}
\label{fat-tree:cable-count}

Cables that interconnect edge and core levels are laid out at installation time, and later updates are difficult and costly.

Therefore, when future use of spare ports is anticipated, we recommend establishing a full fabric between edge and core levels. As we have $E$ edge switches, whose $P_{Ec}$ ports are connected to core level, we need $E \cdot P_{Ec}$ cables to establish a full fabric between the layers. Additionally, $N$ cables are needed to connect compute nodes to the edge level.

The complete expression for the number of cables is therefore $L=N+E \cdot P_{Ec}$. For example, in Figure~\ref{fat-tree:fig:60_nodes_01} the number of cables is $L=60+4 \cdot 18=132$.

For blade servers, the connection of compute nodes to edge switches doesn't require cables, therefore the first summand in the above formula is eliminated.

% Experimental results
\section{Experimental results}
\label{fat-tree:experimental-results}

We apply the proposed algorithm to a real-life scenario, perform detailed calculations and discuss economic implications. Prices are subject to change over time, but this does not affect generality of conclusions.

We build a fat-tree network for a cluster of $N=224$ blade servers. The cluster is built with 14 enclosures, each of them containing 16 blade servers. Every enclosure is fitted with an edge switch with $P_E=32$ ports.

\subsection{Design database}
\label{fat-tree:design-database}

We will use two possible core switches: (a) a 36-port monolithic switch with a price of \$11,000, which is roughly \$306 per port, and (b) a modular switch which can be configured with up to 108 ports, in multiples of 18. In the network design tool this modular switch is represented as six switches of 18, 36, ..., 108 ports. Therefore, $\mathbb{C}$ contains seven items in total.

The modular switch consists of a chassis (\$25,000), 3 fabric boards necessary to make the switch non-blocking (\$9,000 each), and a required number of line cards (up to 6), providing 18 ports each (\$13,000 each). Full configuration costs \$130,000, or \$1,204 per port. This is a four times higher price per port compared to a simple 36-port switch.

Modular switches have a lower port density, therefore using them can unexpectedly increase the total space taken by network equipment. If only limited space is available, this can be dealt with by imposing constraints on equipment size when running the algorithm.

\subsection{Possible core level configurations}
\label{fat-tree:core-level-configs}

According to the algorithm, on the edge level $E=14$ switches will be used. On the core level, there are seven possible choices of core switches. The least expensive configuration of the core level (\$88,000) is obtained when using eight 36-port monolithic switches.

Reduced configurations of the modular switch with 18, 36, 54 and 72 ports result in unreasonably high costs, and we don't analyse them here. They could also be discarded using the following heuristic: modular switches are cost-effective when configured close to their full capacity.

Of special interest are, however, configurations with 90 and 108 ports. Both of them require $C=3$ core switches. The 90-port configuration will be used, as it is slightly cheaper (\$117,000) than the full 108-port configuration (\$130,000). The cost of core level with this configuration is therefore $3 \cdot \$117,000=\$351,000$. This is roughly 4 times more expensive than with a 36-port switch.

\subsection{Factoring in other costs}
\label{fat-tree:factoring-other-costs}

We continue to compare two configurations of the core level -- with 36-port switches and with 90-port switches. Let us now factor in the cost of 14 edge-level switches, located in enclosures (one switch per enclosure, \$11,000 each), and the cost of 224 cables (as per section \ref{fat-tree:cable-count}) of \$80 each (an averaged price for cables of this length, calculated manually). The per-port total costs of the two networks are \$1,160 and \$2,334, respectively -- a twofold difference.

% How it was calculated:
%
% Monolithic switches + edge + cables:
% (88+14*11+224*0,080) / 224 servers = $259,920/224 = $1,160 per port
%
% Modular switches + edge + cables:
% (351+14*11+224*0,080) / 224 servers = $522,920/224 = $2,334 per port

If we further add the cost of blade servers, equipped with dual CPUs, memory and InfiniBand adapters (\$9,600 per each server), and cost of 14 enclosures (\$7,500 per each), we will receive the total costs of the computer cluster, \$2,515,320 and \$2,778,320, for networks made of monolithic and modular core switches, respectively. The difference per connected server diminishes to 10,4\%. It means that for a small premium we can attain a possibility of future network expansion and greatly simplify cabling.

% How it was calculated:
%
% 224 blade servers:
% 9600*224=2,150,400
%
% 14 enclosures:
% 7500*14=105,000
%
% Total: $2,255,400

Additionally, these calculations demonstrate that using blocking networks, such as \textit{thin-trees}, will only marginally reduce total cost of the supercomputer, while potentially having severe consequences on performance.

It is worth noting that cost of \textit{cables} is very low compared to the cost of entire computer cluster. This justifies the use of rough approximations of cable costs when designing entire computers (see also section~\ref{fat-tree:calculating-cable-costs}).

\section{Other design considerations}
\label{fat-tree:other-design-considerations}

\subsection{Per-port metrics}
% Per-port metrics (cost, power consumption)

Let us consider a particular case of a network where edge and core switches are identical, and all ports are occupied. Some useful metrics can be derived for such networks.

Let us denote port count on edge and core switches by $P$. The network can connect a maximum of $N=N_{max}=P^2/2$ nodes. There will be $P$ edge and $P/2$ core switches, for a total of $3P/2$ switches. As switches have $P$ ports each, the total number of ports on all switches will equal to $3P^2/2=3N$, which is thrice the number of nodes. In other words, for each of $N$ connected nodes, the two-layer fat-tree network employs three ports (and a three-layer network employs five).

Several important characteristics, such as network cost and power consumption, are ``additive'' in a sense that $x$ identical switches cost $x$ times more than a single switch and consume $x$ times more power. The same applies on a per-port level: a network of identical arbitrarily connected switches, with a total port count of $y$, costs $y$ times more and consumes $y$ times more power than per-port cost and power consumption, respectively.

This allows to easily determine a rough estimation of cost, power, rack space, weight and possibly other characteristics of a network that supports $N$ nodes by simply multiplying corresponding per-port characteristics of switches by $3N$, without the need for detailed analysis.

For example, a 36-port switch mentioned in section \ref{fat-tree:design-database} has a cost of \$306 per port. Typical power consumption reported by manufacturer is 152W with copper cables, which is 4,22W per port. The switch occupies 1U of rack space, hence ``per-port rack space'' is 1/36.

For a full configuration of $N=648$ nodes, power consumption is $3N$ times per-port consumption (8,204W). Cost is $3N$ times per-port cost (\$594,864). Occupied rack space is $3N$ times ``per-port rack space'' and equals 54U.

This estimation is also correct when the number of nodes $N$ is $X$ times smaller than $N_{max}$, where $X$ is a non-trivial factor of $P/2$. In this case links between core and edge layers run in bundles of $X$. For example, if $P=36$, valid values for $X$ are 2, 3, 6 and 9. The estimation will thus be accurate for clusters of 324, 216, 108 and 72 nodes.

In all other cases there will be spare ports on core and possibly edge layers, and the above approach will systematically underestimate metric values, because the actual network will have more than $3N$ ports. Therefore, the estimation provides the \textit{lower bound} on metric values.

\begin{figure}
\centering
\includegraphics[scale=0.675]{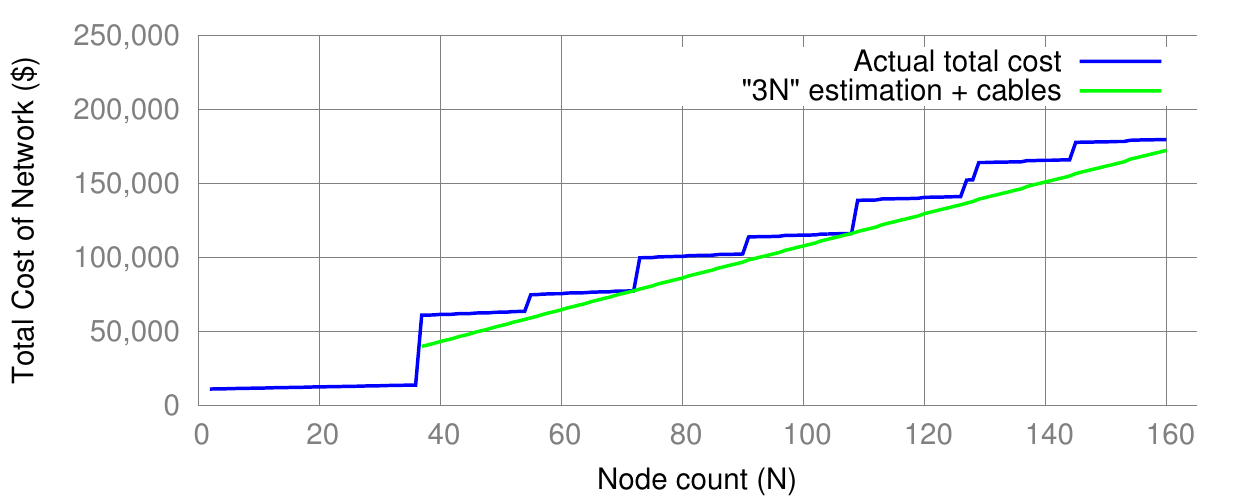}
\caption{Network cost, estimated and actual.}
\label{fat-tree:fig:network_cost_2_162}
\end{figure}

Figure~\ref{fat-tree:fig:network_cost_2_162} provides an example. The blue line represents the actual cost of network built with 36-port switches, including cables, for $N \in [2,160]$ nodes. A region of $2 \le N \le 36$ nodes represents a trivial case of star network, with only one switch: no fat-tree is required, hence the cost of network is kept low. Starting from 37 nodes, a two-layer fat-tree is used. Stepped behaviour of the blue curve is explained by increased switch count for every additional $P/2=18$ nodes. Monotonic increase inside a step is caused by increased cable count for every connected node.

% It was obtained by using web service that designs fat-tree networks.
% Use these commands to query the web service in MS Windows,
% and write the cost of the network into the file:
%
% del fattrees.txt
% for /l %i in (1,1,160) do wget -O - "http://localhost:8000/cgi-bin/network/network.exe?task=design&network_topology=fat-tree&nodes=%i"|awk -F= "/^network_cost/ { print $2 }" >> fattrees.txt

The green line starts at 37-th node and represents the above estimation: $3N$ multiplied by per-port cost, plus the cost of cables. At 72 and 108 nodes it exactly matches the actual cost, as discussed above, but in other points a discrepancy is observed, with the median value of 12\%.

This result allows to quickly obtain engineering evaluations of fat-tree characteristics without referring to the algorithm.

% N=2..36: star network, one switch
% N=37..54: 5 switches total
% N=55..72: 6 switches total
% N=73..90: 8 switches total
% N=91..108: 9 switches total
% N=109..126: 13 switches total, biggest discrepancy from estimation
% N=127..144: 14 switches total
% N=145..162: 15 switches total

\subsection{Designing for future expansion}

Expanding existing fat-tree networks can be a difficult task. While adding edge level switches is easy, core level switches might not have spare ports necessary for expansion if this was not taken care of during design phase. We suggest to design a core level for the largest anticipated number of ports, and then gradually connect more nodes via additional edge switches as the need arises.

\begin{figure}
\centering
\includegraphics[scale=0.4]{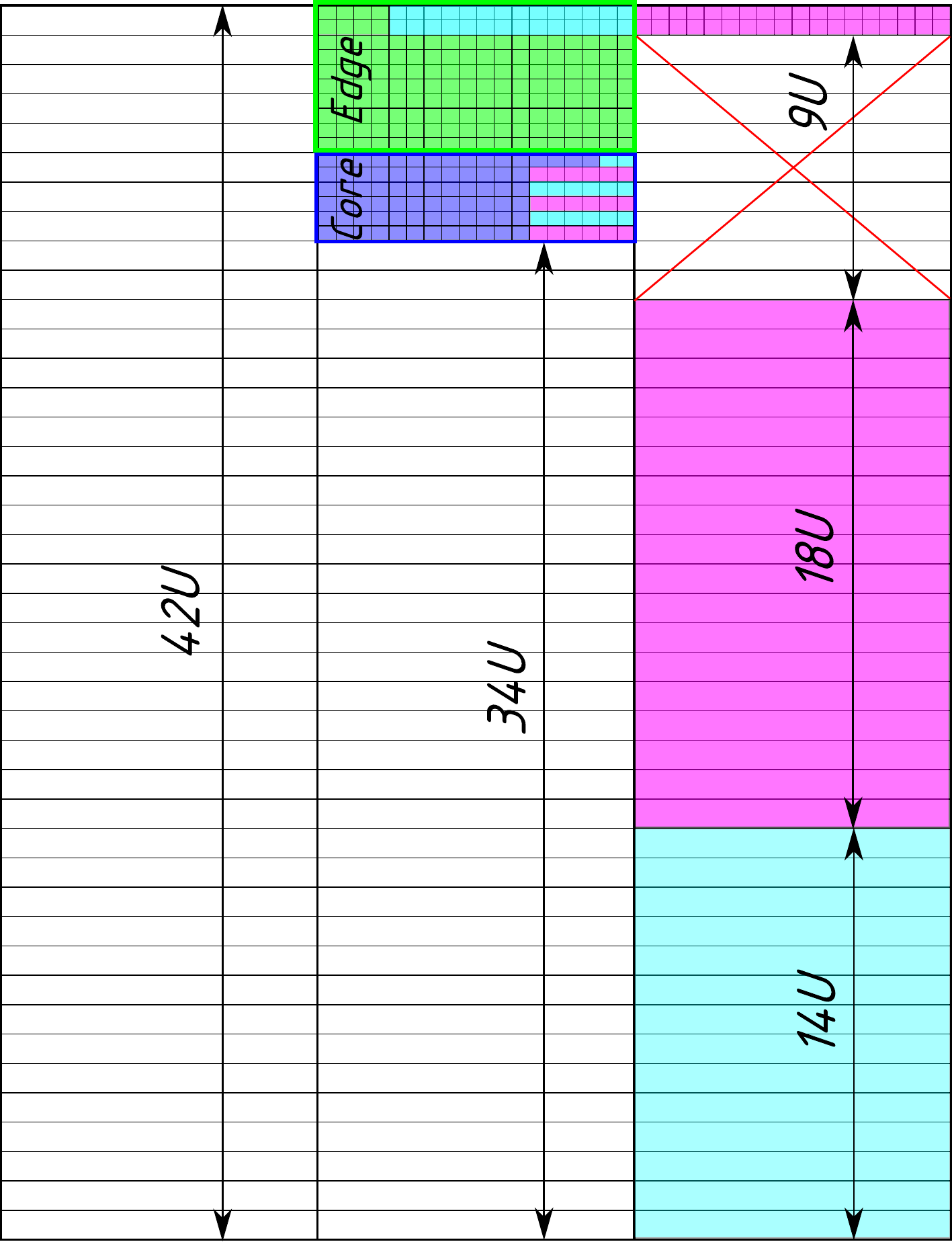}
\caption{Equipment placement for partly expandable configuration.}
\label{fat-tree:fig:future_exp_n_76}
\end{figure}

\begin{figure}
\centering
\includegraphics[scale=1]{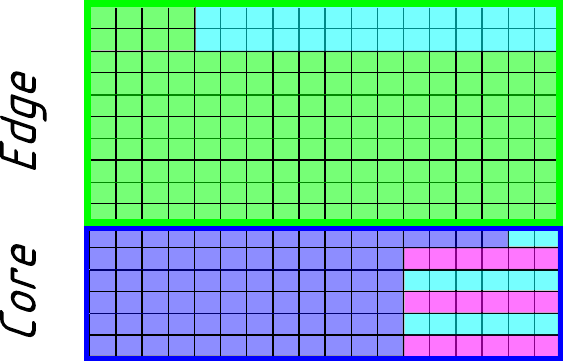}
\caption{Close-up of the area of interest.}
\label{fat-tree:fig:future_exp_n_76_closeup}
\end{figure}

We demonstrate with a real-life scenario that failure to properly construct the core level can lead to non-expandable networks. Suppose we need to build a computer cluster using 1U compute nodes and commodity switches with $P=36$ ports. Additional design constraint is that the cluster shall initially occupy two standard 42U racks with as many nodes as possible, and be expandable to three racks in the future -- imitating lack of space in the machine room.

If we do not take expandability into account, the network design process proceeds as follows. Two racks contain 84U of space. With $N=84$ nodes we require $E=5$ edge and $C=3$ core switches which would occupy additional 8U, thus exceeding allotted space. Hence we reduce node count to $N=76$.

Resulting equipment placement is presented in Figure~\ref{fat-tree:fig:future_exp_n_76}, with a close-up of the area of interest in Figure~\ref{fat-tree:fig:future_exp_n_76_closeup}.

Initially there are two racks, the left and the middle. Five edge switches are located in the top of the middle rack, followed by three core switches. All remaining space is occupied by $N=42+34=76$ compute nodes.

Let us discover opportunities for expandability when the third rack becomes available. First, we can use spare ports in already installed edge and core switches. There are 28 spare ports in the topmost edge switch, shown in cyan, and 32 spare ports in three core switches, shown in cyan and magenta.

Using 28 spare ports in the edge switch allows to connect 14 more nodes which would be placed to the bottom of the newly available right rack, denoted by a cyan block. Of 28 ports, half would be connected to the new nodes, and the remaining half would be connected to corresponding cyan ports of core switches.

Next opportunity for expansion lies in installing a sixth edge switch in the top of the right rack, shown in magenta. 18 ports of this switch will be connected to 18 nodes in the right rack, denoted by a magenta block. Remaining 18 ports would be connected to corresponding magenta ports of core switches.

Now opportunities for expansion are exhausted. A total of $N=108$ nodes were connected, and nine units of rack space cannot be utilized, violating the design constraint. Further expansion would require redesigning the core layer: adding more switches and rewiring connections. For large clusters this is a complex and costly task that should be avoided if possible.

Instead, we can design a network from the start to accommodate the largest anticipated number of nodes. In this case, three racks can house $N=126$ nodes. The network will consist of $E=7$ edge and $C=4$ core switches, occupying in total 11U of rack space. Hence node count shall be respectively reduced to $N=115$. All three racks will be fully populated.

This allows to expand the cluster by 7 additional nodes, compared to the previous variant. However, this also incurs increased network cost, as 11 switches are used instead of 9, so design decisions have to be carefully balanced.

For this newly designed expandable network, there are two alternative variants of equipment installation in the initial two racks:

\begin{enumerate}
\item Install all $C=4$ core and $E=7$ edge switches at once. This requires 11U of racks space, and leaves space for 73 nodes. Additional 42 nodes will be added when a new rack is available.
\item Install all $C=4$ core switches and as many edge switches as required to fill up two racks with nodes, namely, $E=5$ edge switches. This requires 9U of rack space, and allows to install 75 nodes. Additional 40 nodes and two edge switches will be added when a new rack is available.
\end{enumerate}

The latter variant allows to populate initial two racks with more nodes and reduce original investment in edge switches, as their procurement can be delayed until the expansion stage.

\subsection{Placing equipment and calculating cable costs}
\label{fat-tree:calculating-cable-costs}

When designing networks using the proposed algorithm, the \textit{cost} of switches is immediately known, as it follows from the \textit{number} of switches. The other part -- the cost of cables -- is not instantly available, because it depends on cable length which is not known yet.

We saw in section~\ref{fat-tree:factoring-other-costs} that cables represent a small portion of cost of the entire computing system. For draft designs it is beneficial to use a ``typical'' cable cost as an approximation. For final designs, precise cable lengths (which drive procurement decisions) can be obtained after routing phase, although this is a computationally expensive procedure.

Moreover, cable routing depends on such input parameters as rack sizes, server sizes, machine room dimensions and other factors. As a result, routing only becomes possible after positions of equipment in racks and placement of racks on the floor plan becomes known. Therefore we suggest to postpone cable routing until the final stages of the design process.

A straightforward algorithm by Mudigonda \textit{et al.} \cite{mudigonda2011taming} can be employed to perform cable routing. This algorithm relies on a prior placement of nodes and switches into racks in such a manner that minimizes total cable length. Such positioning resembles a knapsack problem, so a number of heuristics were introduced in the cited paper.

As the network design algorithm proposed in our article uses switches of differing sizes, we also propose certain heuristics.

We show below that fat-trees are perfectly suitable for packing racks as densely as possible, or for leaving as much blank space in every rack as desired (e.g., for other equipment), and also allow for a smooth transition between these extreme cases. After running the network design algorithm, the number and types of required switches are obtained, hence the size they occupy becomes available. The same is true for compute nodes. Therefore the number of racks required to house the equipment becomes known. Assuming that machine room dimensions are specified, racks can be placed according to mechanical and cooling requirements. Then the equipment placement stage can begin, using the following heuristics.

\begin{enumerate}

\item Modular switches are physically indivisible and should be placed first. Otherwise one may end up with partially filled racks, with no space in any of them enough to house relatively big modular switches, and new empty racks would be required.

\item Space for other indivisible equipment can be reserved in the same manner.

\item As bundles of cables run from all edge switches to all core switches, the placement of core switches becomes important. They could be placed in the geometrical centre of a machine room, or put uniformly around the room, or otherwise. This requires further investigation.

\item The principal ``building block'' of a fat-tree network is an edge switch and nodes connected to it. It is logical to keep this switch and its nodes in the same rack (with blade servers, this occurs by itself). They are connected with relatively short cables. One or more such building blocks are put in a rack, until one of the following conditions occurs:

\begin{enumerate}
\item The remaining free space in the rack cannot accommodate another building block;
\item Adding more blocks would exceed the weight budget of the rack stipulated by the floor load limit;
\item Adding more blocks would exceed the power consumption of the equipment in the rack above capacity of the power supply or cooling systems
\end{enumerate}

We then calculate the remaining budget of all three characteristics (space, weight and power) for the current rack, estimate the number of compute nodes that can be placed in this rack at a later time, and proceed to the next rack.

\item The previous step -- placing building blocks into racks -- is repeated multiple times, until we have enough semi-filled racks to accommodate the next building block by ``spreading'' it among these racks.

This strategy allows to fill racks as densely as possible, but results in irregular wiring patterns, as noted by other researchers. However, minimizing the number of racks not only saves floorspace, it also allows to decrease the length of cables running between distant racks. If there is no goal to save space, this heuristic could be omitted.

\emph{Example. When using commodity 1U InfiniBand switches with $P=36$ ports and 1U compute nodes, a building block consists (in the case of a non-blocking network) of one switch and 18 nodes, for a total of 19U. Two such blocks can be placed in a standard 42U rack, and 4U of blank space will remain. After filling five racks in this manner, the resultant 20U of space are enough to house the next 19U block.}

\item Although an edge switch and its nodes should preferably be kept in a single rack, they don't necessarily have to be adjacent. In fact, it is recommended to place edge switches as close to the top of rack as allowed by cables that go down to nodes. This decreases the length of cable bundles that run between racks, potentially enabling the use of shorter cables.

\item Racks in a row are filled until the end of a row in a machine room is encountered. The next rack to be filled is chosen across the aisle that separates rows.

This behaviour ensures that parts of the next block to be spread among several racks will remain close to each other.

Placement proceeds in a serpentine pattern, until all equipment is placed. These heuristics are general enough to enable placement of compute nodes and switches of differing physical sizes into racks of differing heights.

\end{enumerate}

After placement is complete, the routing algorithm is run, and cable lengths and costs are obtained. A weighted average of cable prices can then be used as an average cable price for a cluster of that size (a bigger cluster will need more racks and thus will have a greater average cable length). Throughout this article, an average cable price of \$80 was used, obtained using a similar procedure.

\begin{figure}
\centering
\includegraphics[scale=0.4]{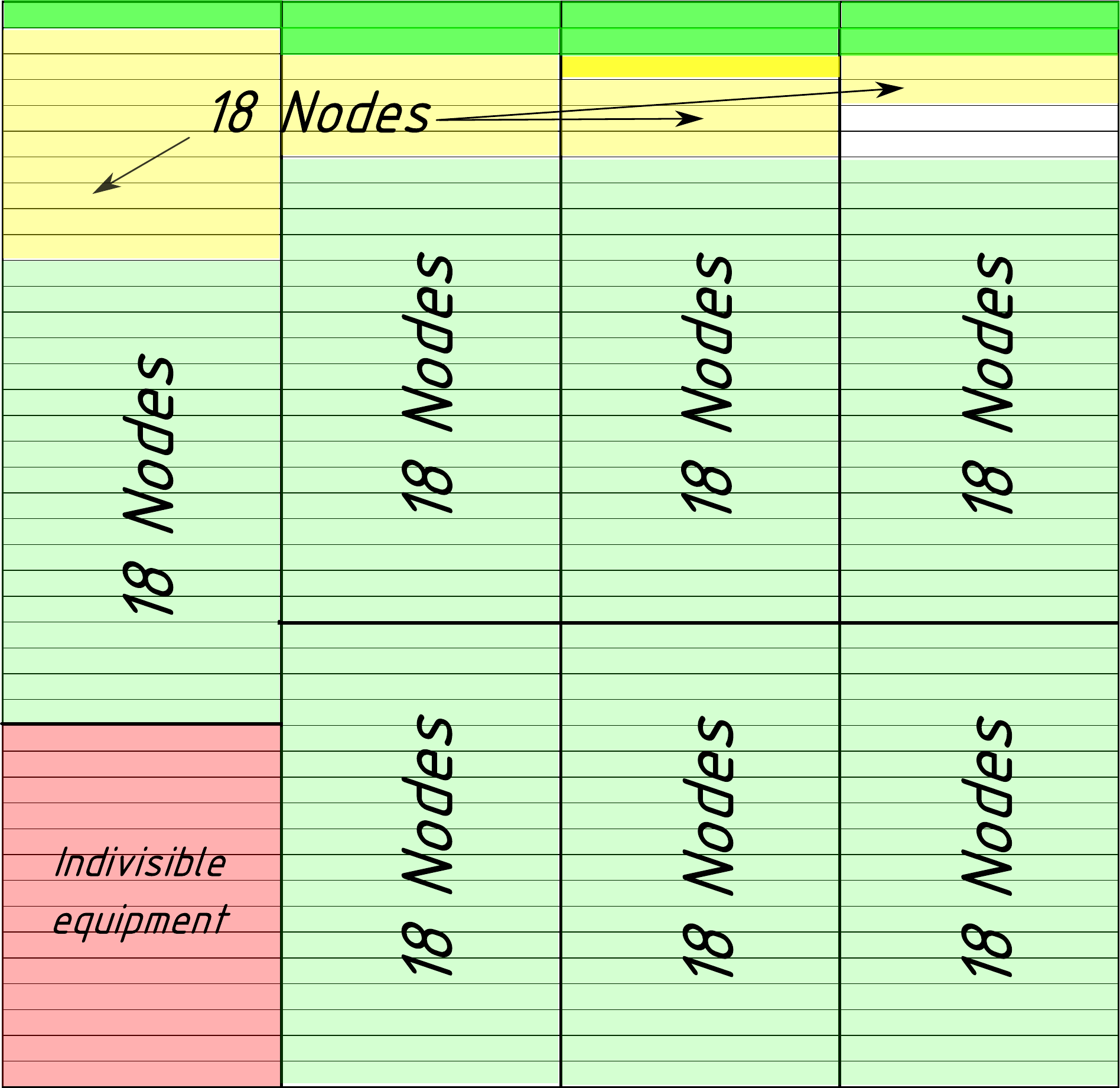}
\caption{Front view of four racks filled using the proposed heuristic (see description in text).}
\label{fat-tree:fig:placement_front_view}
\end{figure}

We present an example of using this heuristic on Figure~\ref{fat-tree:fig:placement_front_view}, deliberately demonstrating filling the racks as densely as possible. In this example, we need to design a non-blocking fat-tree network for $N=396$ compute nodes, using switches with $P=36$ ports. This configuration requires $E=22$ edge and $C=18$ core switches.

\begin{figure}
\centering
\includegraphics[scale=0.55]{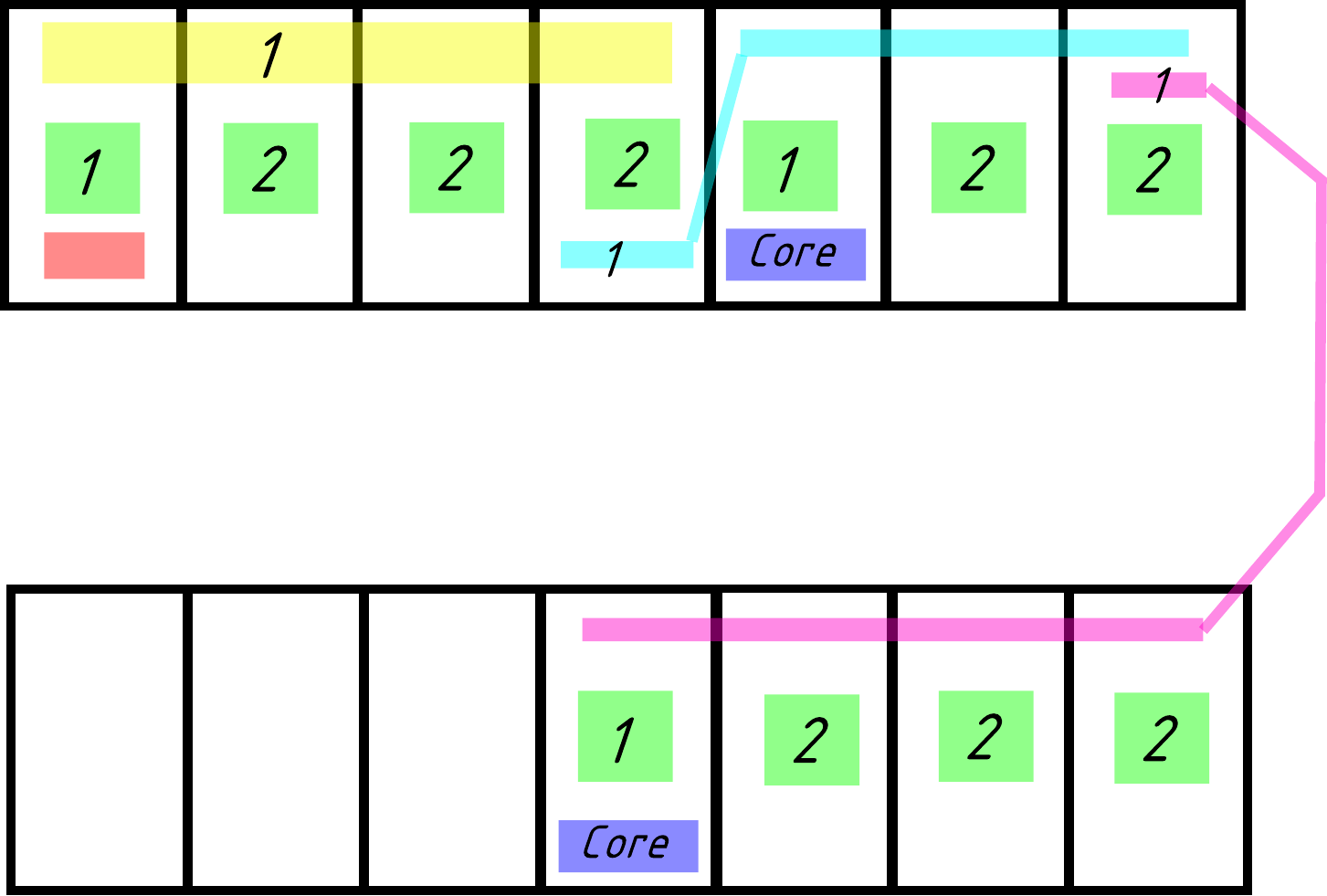}
\caption{Top view of 14 racks filled using the proposed heuristic (see description in text).}
\label{fat-tree:fig:placement_top_view}
\end{figure}

The large indivisible blocks of equipment are installed first (red). Then, we install blocks of compute nodes (pale green), and edge switches for the blocks are put to the top of corresponding racks (bright green). Finally, after installing seven such blocks into four racks, we have enough spare space to place one more block of compute nodes (pale yellow), however, this time it has to be ``spread'' among all four racks. The exact location of the corresponding edge switch for this block (bright yellow) can be chosen arbitrarily. Two units of space in the rightmost rack are empty and will be used further.

The top view of 14 racks, arranged in two rows, is presented on Figure~\ref{fat-tree:fig:placement_top_view}. First four racks from the upper row correspond to those on Figure~\ref{fat-tree:fig:placement_front_view}. After the indivisible equipment is placed, we place 19 out of 22 blocks of compute nodes and their edge switches (green). (The 18 core switches are placed arbitrarily in this example, in two blocks, denoted by violet colour).

By now, 11 racks have been occupied, and there is enough spare space left to accommodate the remaining three blocks of compute nodes, which will have to be ``spread'' in the spare space. The first two of them, denoted by yellow and cyan, are spread along four racks. The last block, denoted by magenta, spreads along five racks and crosses the aisle that separates rows. This dense placement approach is not very elegant; however, in this particular example it allows to use 11 racks instead of 12.

\begin{comment}
BTW, modular switches take more space - their port density per unit is about 60\% lower.
\end{comment}

% Conclusions and future work
\section{Conclusions and future work}
\label{fat-tree:conclusions}

We present the algorithm to automatically design two-layer fat-tree networks with arbitrary blocking factors. We apply proposed algorithm to design several networks and analyse their characteristics. We demonstrate that a lower bound (and a rough approximation) for many technical and economic characteristics of the whole network can be easily obtained from per-port metrics. We also discuss issues of expandability of fat-trees and propose placement heuristics.

Directions of future research include further exploration of reduced tree topologies \cite{navaridas2010reducing,kamil2010communication} for large supercomputers and estimating savings in capital and operating expenditure.

\section{Acknowledgements}
The author is grateful to Dr.~Jos\'e Duato for his insightful comments on the draft of this article.

% !!!
%
% Insert this command somewhere on the last page:
%\balancecolumns 

\bibliographystyle{abbrv}
\bibliography{../../text/bibliography/literature}

\end{document}